\documentclass[12pt,journal,onecolumn]{IEEEtran}
\usepackage{amsmath,amssymb,threeparttable,placeins,makecell,stfloats,cite,booktabs}
\usepackage[dvips]{graphicx,epstopdf}
\usepackage[usenames]{color}
\usepackage{amsfonts}
\usepackage{latexsym}
\usepackage{setspace}
\usepackage{caption}
\usepackage{subcaption}
\usepackage{floatrow}
\newtheorem{theorem}{{{\textit{Theorem}}}}

\newtheorem{lemma}{{{\textit{Lemma}}}}

\newtheorem{property}{{{\textit{Property}}}}
\newtheorem{definition}{{{\textit{Definition}}}}

\newtheorem{remark}{{{\textit{Remark}}}}

\newtheorem{example}{{{\textit{Example}}}}

\renewcommand{\arraystretch}{1.2}

\evensidemargin=\oddsidemargin
\begin{document}
\doublespacing{}
\title{Asymptotically Optimal Golay-ZCZ Sequence Sets with Flexible Length}
\author{Zhi Gu,~\IEEEmembership{Student Member,~IEEE,}
Zhengchun Zhou,~\IEEEmembership{Member,~IEEE,}
    Avik Ranjan Adhikary,~\IEEEmembership{Member,~IEEE,}
    Yanghe Feng,~
    Pingzhi Fan,~\IEEEmembership{Fellow,~IEEE.} 
\thanks{Z. Gu is with the School of Information Science and Technology, Southwest Jiaotong University, Chengdu, 611756, China. E-mail: goods@my.swjtu.edu.cn.}
\thanks{Z. Zhou and A. R. Adhikary are with the School of Mathematics, Southwest Jiaotong University,
Chengdu, 611756, China. E-mail: zzc@swjtu.edu.cn, Avik.Adhikary@ieee.org.}
\thanks{Y. Feng is with the College of Systems Engineering, National University of Defense Technology,
Changsha, 410073, China. E-mail: fengyanghe@nudt.edu.cn}
\thanks{P. Fan is with the Institute of Mobile Communications, Southwest Jiaotong University, Chengdu, China. E-mail: pzfan@swjtu.edu.cn}
\thanks{The material in this
	paper was presented in part at the 2021 IEEE International Symposium on
	Information Theory.}
}

\maketitle

\date{}

\begin{abstract}
Zero correlation zone (ZCZ) sequences and Golay complementary sequences are two kinds of sequences with different preferable correlation properties. Golay-ZCZ sequences are special kinds of complementary sequences which also possess a large ZCZ and are good candidates for pilots in OFDM systems. Known Golay-ZCZ sequences reported in the literature have a limitation in the length which is the form of a power of 2. In this paper, we propose two constructions of Golay-ZCZ sequence sets with new parameters which generalize the constructions of Gong et al. (IEEE Transaction on Communications 61(9), 2013) and  Chen et al (IEEE Transaction on Communications 61(9), 2018). Notably, one of the constructions results in optimal binary Golay-ZCZ sequences, while the other results in asymptotically optimal polyphase Golay-ZCZ sequences as the number of sequences increases.

%
\end{abstract}


\section{Introduction}\label{section 1}
Golay complementary sets (GCS) and zero correlation zone (ZCZ) sequence sets are two kinds of sequence sets with different desirable correlation properties. GCS are sequence sets have zero aperiodic autocorrelation sums (AACS) at all non-zero time shifts \cite{Golay61}, whereas ZCZ sequence sets have zero correlation zone within a certain time-shift \cite{Fan2007}. Due to its favourable correlation properties GCSs or ZCZ sequence sets have been widely used to reduce peak average power ratio (PAPR) in orthogonal frequency division multiplexing systems \cite{Davis1999,Paterson2000}. However, the sequences own periodic autocorrelation plays an important role in some applications like synchronization and detection of the signal. 

Working in this direction, Gong \textit{et al.} \cite{Gong2013} investigated the periodic autocorrelation behaviour of a single Golay sequence in 2013. To be more specific, Gong \textit{et al.} presented two constructions of Golay sequences of length $2^m$, using generalized Boolean functions (GBF), each displaying a periodic zero autocorrelation zone (ZACZ) of $2^{m-2}$, and $2^{m-3}$, respectively, around the in-phase position \cite{Gong2013}. In 2018, Chen \textit{et al.} studied the zero cross-correlation zone among the Golay sequences and proposed Golay-ZCZ sequence sets \cite{Chen201811}. Golay-ZCZ sequence sets are sequence sets having periodic ZACZ for each sequences, periodic zero cross-correlation zone (ZCCZ) for any two sequences and also the aperiodic autocorrelation sum is zero for all non-zero time shifts. Specifically, using GBFs, Chen \textit{et al.} gave a systematic construction of Golay-ZCZ sequence set consisting $2^k$ sequences, each of length $2^m$ and $\min\{ZACZ,ZCCZ\}$ is $2^{m-k-1}$ \cite{Chen201811}.

In \cite{Gong2013}, the authors discussed the application of Golay sequences with large ZACZ for ISI channel estimation. Using Golay sequences with large ZACZ as channel estimation sequences (CES), the authors analysed the performance of Golay-sequence-aided channel estimation in terms of the error variance and the classical Cramer-Rao lower bound (CRLB). It was shown in \cite{Gong2013} that when the channel impulse response (CIR) is within the ZACZ width then the variance of the Golay sequences attains the CRLB. Recently in 2021, Yu \cite{yu} demonstrated that sequence sets having low coherence of the spreading matrix along with low PAPR are suitable as pilot sequences for uplink grant-free non-orthogonal multiple access (NOMA). The work of \cite{yu} depicts that Golay-ZCZ sequences can be suitably used as pilot sequences for uplink grant-free NOMA.

Inspired by the works of Gong \textit{et al.} \cite{Gong2013} and Chen \textit{et al.} \cite{Chen201811} and by the applicability of the Golay-ZCZ sequences as pilot sequences for uplink grant-free NOMA and channel estimation, we propose Golay-ZCZ sequence sets with new lengths. Note that, the lengths of the GCPs with large ZCZs discussed in the works of Gong \textit{et al.} \cite{Gong2013} and Chen \textit{et al.} \cite{Chen201811} are all in the powers of two. To the best of the authors knowledge, the problem of investigating the individual periodic autocorrelations of the GCPs and the periodic cross-correlations of the pairs when the length of the GCPs are non-power-of-two, remains largely open. An overview of the previous works, which considers the periodic ZACZ of the individual sequences and ZCCZ of a GCP, is given in Table \ref{Table duibi}.
\begin{table}
	\caption{Golay sequences with periodic ZACZ and ZCCZ.}
	\label{Table duibi}
	\begin{tabular}{|c|c|c|c|c|c|}
		\hline
		Ref. & Length & set size & ZACZ width & ZCCZ width &Based on \\\hline
		
		\hline
		
		\cite{Gong2013} & $2^m$ & $2$ & $2^{m-2}$ or $2^{m-3}$ & Not discussed & GBF \\
		\hline
		
		\cite{Chen201811} & $2^m$ & $2^k$ & $2^{m-k-1}$ & $2^{m-k-1}$ & GBF \\
		\hline
		
		Theorem 1 & $4N$ & $2$ & $N$ & $N$ & GCP of length $N$. \\
		\hline
		Theorem 2 & $M^2N$ & $M$ & $(M-1)N$ & $(M-1)N$ & $(M,M,N)$- CCC.\\
		\hline
	\end{tabular}     
\end{table}
We have proposed two constructions which results to Golay-ZCZ sequence sets consisting of sequences with more flexible lengths. To be more specific, assuming that Golay complementary pairs of length $N$ exists, we have proposed a systematic construction of Golay-ZCZ sequences of length $4N$, having ZCZ width of $N+1$. One of the limitations of construction of Golay-ZCZ sequence sets reported in \cite{Chen201811} is that, the ZCZ width decreases, as the number of sequences increases. To solve that problem, we proposed another construction of Golay-ZCZ sequences consisting of sequences of length $M^2N$ having ZCZ length $(M-1)N$ from an $(M,M,N)$ compete complementary code (CCC). Interestingly, the resultant Golay-ZCZ sequences derived from the CCC are asymptotically optimal with respect to Tang-Fan-Matsufuji bound \cite{tangfan} as the number of sequences increases, when polyphase sequences are considered. To increase the availability of the CCCs to design Golay-ZCZ sequences of more flexible lengths we have also proposed a new iterative construction of CCCs based on Kronecker product. A brief overview of all the parameters of CCCs proposed till date can be found in Table \ref{tabccc}. 
\begin{table}
	\centering
	\resizebox{\textwidth}{!}{
	\caption{Parameters of CCCs}
	\label{tabccc}
	\begin{tabular}{|c|c|c|c|}
		\hline
		Ref. & Parameters & constraints & Construction based on   \\\hline
		
		\hline
		
		\cite{suherio15} & $(M,M,M^N)$ & $N\geq 2$ & Unitary matrices  \\
		\hline
		\cite{huang18} & $(2^{N-r},2^{N-r},2^N)$ & $r=1,2,\dots,N-1$ & Golay-paired Hadamard matrices  \\
		\hline
		\cite{han53} & $(M,M,MN)$ & $N\leq M$ & Unitary matrices  \\
		\hline
		\cite{han17} & $(M,M,MN/P)$ & $N,P\leq M$ & Unitary matrices  \\
		\hline
		\cite{zhang54} & $(M,M,2^mM)$ & $m\geq 1$ & Unitary matrices  \\
		\hline
		\cite{rathinakumar21} & $(2^{k+1},2^{k+1},2^m)$ & $m,k\geq 1,~k=m-1$ & GBF  \\
		\hline
		\cite{yang55} & $(2N,2N,l)$ & $l> 1,~N\geq 1$ & Unitary matrices   \\
		\hline
		\cite{mar22} & $(2^m,2^m,2^{mN})$ & $m\geq 1$ & Equivalent Hadamard matrices   \\
		\hline
		\cite{chen20} & $(2^m,2^m,2^k)$ & $m,k\geq 1$ with $k\geq m$ & GBF   \\
		\hline
		\cite{liu19} & $(2^m,2^m,2^k)$ & $m,k\geq 1$ & GBF   \\
		\hline
		\cite{das23,ma} & $(M,M,M^N)$ & $N\geq 1$ & Paraunitary (PU) matrices   \\
		\hline
		\cite{das24} & $(M,M,P^N)$ & $N\geq 1,~P|M$ & PU matrices   \\
		\hline
       \cite{jin} & $(M_1M_2,M_1M_2,N_1N_2)$ & $(M_1,M_1,N_1)$- CCC and $(M_2,M_2,N_2)$- CCC exists & Kronecker product   \\
\hline
Proposed & $(M,M,MN_1N_2)$ & $(M,M,N_1)$- CCC and $(M,M,N_2)$- CCC exists & Kronecker product   \\
\hline
	\end{tabular}}
\end{table}

The rest of the paper is organized as follows. In Section \ref{section 2}, some useful notations and preliminaries are recalled.
In Section \ref{section 3}, a systematic construction of Golay-ZCZ sequence pairs with flexible lengths is proposed. In Section \ref{golayzcz2d}, a systematic construction of Golay-ZCZ sequence sets is proposed based on existing CCCs. Also in this section, we have proposed an iterative construction of CCCs to increase the flexibility of the parameters of the proposed Golay-ZCZ sequences. In Section \ref{secv}, we have discussed the optimality of the proposed Golay-ZCZ sequence sets. In Section \ref{novel}, we have discussed the novelty of the proposed constructions as compared to previous works. Finally, we conclude the paper in Section \ref{section 5}.

\section{Preliminaries}\label{section 2}
Before we begin, let us define the following notations:
\begin{itemize}
	\item $\mathbf{0}_L$ denotes the all-zero vector of length-$L$.
	\item $\overleftarrow{\mathbf{a}}$ denotes the reverse of the sequence $\mathbf{a}$.
	\item $x^*$ denotes the complex conjugate of $x$.
	\item $\mathbf{a}||\mathbf{b}$ denotes the concatenation of the sequences $\mathbf{a}$ and $\mathbf{b}$.
	\item `$\mathbf{a}\cdot \mathbf{b}$' denotes the `inner product' of two sequences $\mathbf{a}$ and $\mathbf{b}$.
	\item $<x>_M$ denotes $x \mod M$.
	\item $\mathbf{x}\otimes\mathbf{y}= [x_0\mathbf{y},x_1\mathbf{y},\cdots, x_{N_1-1}\mathbf{y}]$, denotes the Kronecker product of the sequences $\mathbf{x}$ and $\mathbf{y}$.
\end{itemize}
\begin{definition}
	Let $\mathbf{a}$ and $\mathbf{b}$ be two length $N$ sequences.
	The periodic cross-correlation function (PCCF) of ${\mathbf{a}}$ and ${\mathbf{b}}$ is defined as
	\begin{equation}\label{defi_PCCF}
		R_{\mathbf{a},\mathbf{b}}(\tau):= \left \{
		\begin{array}{cl}
			\sum\limits_{k=0}^{N-1}{a_kb^*_{<k+\tau>_N}},&~~0\leq \tau \leq N-1;\\
			\sum\limits_{k=0}^{N-1}{a_{<k-\tau>_N}b^*_k},&~~-(N-1)\leq \tau \leq -1;
		\end{array}
		\right .
	\end{equation}
	When $\mathbf{a} = \mathbf{b}$, $R_{\mathbf{a},\mathbf{b}}(\tau)$ is called periodic auto-correlation function (PACF) of $\mathbf{a}$ and is denoted as $R_{\mathbf{a}}(\tau)$.
\end{definition}

\begin{definition}
	Let $\mathbf{a}$ and $\mathbf{b}$ be two length $N$ sequences.
	The aperiodic cross-correlation function (ACCF) of ${\mathbf{a}}$ and ${\mathbf{b}}$ is defined as
	\begin{equation}\label{defi_ACCF}
		C_{\mathbf{a},\mathbf{b}}(\tau):= \left \{
		\begin{array}{cl}
			\sum\limits_{k=0}^{N-1-\tau}{a_kb^*_{k+\tau}},&~~0\leq \tau \leq N-1;\\
			\sum\limits_{k=0}^{N-1+\tau}{a_{k-\tau}b^*_k},&~~-(N-1)\leq \tau \leq -1;
		\end{array}
		\right .
	\end{equation}
	When $\mathbf{a} = \mathbf{b}$, $C_{\mathbf{a},\mathbf{b}}(\tau)$ is called aperiodic auto-correlation function (AACF) of $\mathbf{a}$ and is denoted as $C_{\mathbf{a}}(\tau)$.
\end{definition}

\begin{definition}
	A set $\mathcal{C} = \{\mathbf{c}_0, \mathbf{c}_1,..., \mathbf{c}_{M-1}\}$ consisting
	of $M$ sequences of length $N$ is said to be an $(M,N,Z)$- ZCZ sequence set if it satisfies 
	\begin{equation}
		\begin{split}
			&R_{\mathbf{c}_i}(\tau)=0,~\text{for }1\leq |\tau|\leq Z,~0\leq i \leq M-1,\\
			&R_{\mathbf{c}_i,\mathbf{c}_j}(\tau)=0,~\text{for }|\tau|\leq Z,~0\leq i\neq j\leq M-1.
		\end{split}
	\end{equation}
\end{definition}

\begin{definition}\label{golayzcz1d}
	An $(M,N,Z)$- ZCZ sequence set becomes an $(M,N,Z)$- Golay-ZCZ sequence set if it satisfies 
	\begin{equation}
		\begin{split}
			&\text{C1: }\sum_{i=0}^{M-1}C_{\mathbf{c}_i}(\tau)=0,~\text{for all }\tau\neq 0,\\
			&\text{C2: }R_{\mathbf{c}_i}(\tau)=0,~\text{for }1\leq |\tau|\leq Z,~0\leq i \leq M-1,\\
			&\text{C3: }R_{\mathbf{c}_i,\mathbf{c}_j}(\tau)=0,~\text{for }|\tau|\leq Z,~0\leq i\neq j\leq M-1.
		\end{split}
	\end{equation}
\end{definition}

\begin{definition}
	Let $\mathcal{C}$ be a $P \times N$ matrix, consisting of $P$ sequences of length $N$, as follows:
	\begin{equation}
		\mathcal{C}=\begin{bmatrix}
			\mathbf{c}_0\\
			\mathbf{c}_1\\
			\vdots\\
			\mathbf{c}_{P-1}
		\end{bmatrix}_{P\times N}=\begin{bmatrix}
		c_{0,0} & c_{0,1} & \dots & c_{0,N-1}\\
		c_{1,0} & c_{1,1} & \dots & c_{1,N-1}\\
		\vdots & \vdots & \ddots & \vdots\\
		c_{P-1,0} & c_{P-1,1} & \dots & c_{P-1,N-1}
	\end{bmatrix}_{P\times N}.
	\end{equation}
	Then $\mathcal{C}$ is called a CS of size $P$ if
	\begin{equation}
		C_{\mathbf{c}_0}(\tau)+C_{\mathbf{c}_1}(\tau)+\cdots+C_{\mathbf{c}_{P-1}}(\tau)=\begin{cases}
			PN & \text{if } \tau=0,\\
			0 & \text{if } 0<\tau<N.
		\end{cases}
	\end{equation}
\end{definition}

\begin{definition}
	Consider $\mathfrak{C}=\left\{\mathcal{C}^{0}, \mathcal{C}^{1}, \cdots, \mathcal{C}^{M-1}\right\}$, consisting $M$ CSs $\mathcal{C}^{k}$, $0\leq k<M$, each having $M$ sequences of length $N$, i.e.,
	\begin{equation}\label{eq2}
		\mathcal{C}^{k}=\begin{bmatrix}
			\mathbf{c}^{k}_0\\
			\mathbf{c}^{k}_1\\
			\vdots\\
			\mathbf{c}^{k}_{M-1}
		\end{bmatrix}_{M \times N}, 0 \leq k \leq M-1,
	\end{equation}
	where $ \mathbf{c}_{m}^{k}$ is the $m$-th sequence of length $N$ and is expressed as $ \mathbf{c}_{m}^{k}=\left(c_{m, 0}^{k}, c_{m, 1}^{k}, \cdots, c_{m, N-1}^{k}\right)$, $0 \leq m \leq M-1$. The set $\mathfrak{C}$ is called a $(M,M,N)$ complete complementary code (CCC) if for any $\mathcal{C}^{k_1},\mathcal{C}^{k_2}\in \mathfrak{C}$, $0\leq k_1,k_2
	\leq K-1$,
	$0 \leq \tau \leq N-1, k_{1} = k_{2}$ or $0<\tau \leq N-1, k_{1}\neq k_{2}$,
	\begin{equation}
		|C_{\mathcal{C}^{k_1},\mathcal{C}^{k_2}}(\tau)|=\left|\sum_{m=0}^{M-1} C_{\mathbf{c}_{m}^{k_{1}},
			{\mathbf{c}_{m}^{k_{2}}}}(\tau)\right|= 0,
	\end{equation}
	where $M$ denotes the set size and the number of sequences in each sequence set, and $N$ the length of constituent sequences of $\mathfrak{C}$. 
\end{definition}

Let $\mathcal{C}$ be an $M\times N$ matrix given by
\begin{equation}
	\mathcal{C}=\begin{bmatrix}
		c_{0,0} & c_{0,1} & \dots & c_{0,N-1}\\
		c_{1,0} & c_{1,1} & \dots & c_{1,N-1}\\
		\vdots & \vdots & \ddots & \vdots \\
		c_{M-1,0} & c_{M-1,1} & \dots & c_{M-1,N-1}
	\end{bmatrix}_{M\times N}.
\end{equation}
Then let us define a transformation $L^{r}_s(\mathcal{C})$ as follows:
\begin{equation}\label{eq12n}
	\begin{split}
		L^{r}_s(\mathcal{C})&=\begin{bmatrix}
		\textcolor{blue}{c_{r,s} }&\textcolor{blue}{ c_{r,s+1} }&\textcolor{blue}{ \dots} &\textcolor{blue}{ c_{r,N-1}} &\textcolor{blue}{ c_{r+1,0}} &\textcolor{blue}{ c_{r+1,1}} &\textcolor{blue}{ \dots} &\textcolor{blue}{ c_{r+1,s-1}}\\
		\textcolor{blue}{	c_{r+1,s} }& \textcolor{blue}{c_{r+1,s+1}} &\textcolor{blue}{ \dots} &\textcolor{blue}{ c_{r+1,N-1}}&\textcolor{blue}{c_{r+2,0}} & \textcolor{blue}{c_{r+2,1}} & \textcolor{blue}{\dots }&\textcolor{blue}{ c_{r+2,s-1}}\\
		\textcolor{blue}{	\vdots} &\textcolor{blue}{ \vdots} &\textcolor{blue}{ \ddots} &\textcolor{blue}{ \vdots} &\textcolor{blue}{ \vdots} & \textcolor{blue}{\vdots} &\textcolor{blue}{ \ddots} &\textcolor{blue}{ \vdots} \\
		\textcolor{blue}{	c_{M-2,s}} &\textcolor{blue}{ c_{M-2,s+1}} &\textcolor{blue}{ \dots} & \textcolor{blue}{c_{M-2,N-1}} &\textcolor{blue}{ c_{M-1,0}} &\textcolor{blue}{ c_{M-1,1}} & \textcolor{blue}{\dots} &\textcolor{blue}{ c_{M-1,s-1}}\\
		\textcolor{blue}{	c_{M-1,s}} & \textcolor{blue}{c_{M-1,s+1}} &\textcolor{blue}{ \dots} & \textcolor{blue}{c_{M-1,N-1}} &\textcolor{red}{ c_{0,0} }&\textcolor{red}{ c_{0,1} }&\textcolor{red}{ \dots} &\textcolor{red}{ c_{0,s-1}}\\
		\textcolor{red}{	c_{0,s}} & \textcolor{red}{c_{0,s+1}} & \textcolor{red}{\dots} &\textcolor{red}{ c_{0,N-1}} & \textcolor{red}{c_{1,0}} &\textcolor{red}{ c_{1,1}} &\textcolor{red}{ \dots} & \textcolor{red}{c_{1,s-1}}\\
		\textcolor{red}{	\vdots} &\textcolor{red}{ \vdots} &\textcolor{red}{ \ddots} &\textcolor{red}{ \vdots} & \textcolor{red}{\vdots} &\textcolor{red}{ \vdots} & \textcolor{red}{\ddots} &\textcolor{red}{ \vdots} \\
		\textcolor{red}{	c_{r-1,s}} &\textcolor{red}{ c_{r-1,s+1}} &\textcolor{red}{ \dots} &\textcolor{red}{ c_{r-1,N-1}} &\textcolor{red}{ c_{r,0}} &\textcolor{red}{ c_{r,1}} &\textcolor{red}{ \dots }&\textcolor{red}{ c_{r,s-1}}
		\end{bmatrix}_{M\times N}\\
	&=\begin{bmatrix}
		T^r(\mathbf{c}_{:,s}) & T^r(\mathbf{c}_{:,s+1}) & \dots & T^r(\mathbf{c}_{:,N-1}) & T^{r+1}(\mathbf{c}_{:,0}) & T^{r+1}(\mathbf{c}_{:,1}) & \dots & T^{r+1}(\mathbf{c}_{:,s-1})
	\end{bmatrix},
	\end{split}
\end{equation}
where $\mathbf{c}_{:,s}$ denotes the $s$-th column of $\mathcal{C}$, and $T^r(\mathbf{c}_{:,s})$ denotes the $r$-step up-shift operator.

\begin{definition}
	Let $\mathbf{a}=(a_0,a_1,\dots,a_{N-1})$ be a sequence of length $N$, then the characteristic polynomial of $\mathbf{a}$ is defined as 
	\begin{equation}
		\mathbf{a}(x)=a_0+a_1x+\cdots+a_{N-1}x^{N-1},
	\end{equation}
and the corresponding complex conjugate is given by
\begin{equation}
	\mathbf{a}^*(x)=a^*_0+a^*_1x+\cdots+a^*_{N-1}x^{N-1}.
\end{equation}
\end{definition}

\begin{definition}
	Let $\mathbf{a}(x)$ and $\mathbf{b}(x)$ be the characteristic polynomial of the length $N$ sequence $\mathbf{a}$ and $\mathbf{b}$, respectively. Then the ACCF is defined as
	\begin{equation}
		C_{\mathbf{a},\mathbf{b}}(x)=\sum_{\tau=0}^{N-1}C_{\mathbf{a},\mathbf{b}}(\tau)x^\tau=\mathbf{a}(x^{-1})\mathbf{b}^*(x).
	\end{equation}
\end{definition}

\begin{definition}
	A pair of sequences $(\mathbf{a}, \mathbf{b})$ is said to be a GCP of length $N$ if 
	\begin{equation}
		C_{\mathbf{a}}(x)+C_{\mathbf{b}}(x)=N.
	\end{equation}
\end{definition}

\begin{definition}
	Consider $\mathfrak{C}=\left\{\mathcal{C}^{0}, \mathcal{C}^{1}, \cdots, \mathcal{C}^{M-1}\right\}$, consisting $M$ CSs $\mathcal{C}^{k}$, $0\leq k<M$, each having $M$ sequences of length $N$, as described in (\ref{eq2}). Then $\mathfrak{C}$ is a CCC if 
	\begin{equation}
		C_{\mathcal{C}^i,\mathcal{C}^k}(x)=\sum_{j=0}^{M-1}C_{\mathbf{c}^i_j,\mathbf{c}^k_j}(x)=MN \delta(i-k),
	\end{equation}
where $\delta$ is Kronecker delta function.
\end{definition}

\section{New Construction of Golay Complementary Pair with ZCZ}\label{section 3}
In \cite{Gong2013} and \cite{Chen201811}, the authors considered complementary sequences whose length is only of the form $2^m$. In this section, we considered GCPs of more flexible lengths and proposed Golay-ZCZ sequences of new lengths. Before introducing the construction, we need the following lemma.


\begin{lemma}\label{lem2}
	Let $(\mathbf{a,b})$ be a GCP of lenth $N$ and $(\mathbf{c,d})$ be its complementary mate. Then
	\begin{equation}
		C^*_{\mathbf{a},\mathbf{b}}(\tau)+ C^*_{\mathbf{c},\mathbf{d}}(\tau) = 0
	\end{equation}
\end{lemma}
\begin{IEEEproof}
	From the properties of aperiodic cross-correlation, for any sequence $\mathbf{x}$ and $\mathbf{y}$, we have $C_{\mathbf{x}, \mathbf{y}}(\tau)=C_{\overleftarrow{\mathbf{y}^*}, \overleftarrow{\mathbf{x}^*}}(\tau)$ and $C_{-\mathbf{x}, \mathbf{y}}(\tau)=-C_{\mathbf{x}, \mathbf{y}}(\tau)$. Since $(\mathbf{c}, \mathbf{d})=(\overleftarrow{\mathbf{b}^*},-\overleftarrow{\mathbf{a}^*})$, therefore $C_{\mathbf{c}, \mathbf{d}}=C_{\overleftarrow{\mathbf{d}^*}, \overleftarrow{\mathbf{c}^*}}(\tau)=C_{-{\mathbf{a}},\mathbf{b}}(\tau)=-C_{\mathbf{a},\mathbf{b}}(\tau)$. Hence the proof follows.
\end{IEEEproof}

\begin{theorem}\label{th1}
	Let $(\mathbf{a,b})$ be a GCP of size $N$ and $(\mathbf{c,d})$ be a Golay complementary mate of $(\mathbf{a,b})$. Define
	\begin{equation}
		\begin{split}
			&\mathbf{p}=
			\left(
			\begin{array}{rrrr}
				x_1\mathbf{a}~ || ~x_2\mathbf{b} ~||~ x_3\mathbf{a} ~||~ x_4\mathbf{b} \\
			\end{array}
			\right),\\&
			\mathbf{q}=
			\left(
			\begin{array}{rrrr}
				x_1\mathbf{c} ~ || ~ x_2\mathbf{d} ~ || ~ x_3\mathbf{c} ~ || ~ x_4\mathbf{d} \\
			\end{array}
			\right),
		\end{split}
	\end{equation}
	where $x_1,x_2,x_3,x_4\in\{+1,-1\}$. Then $(\mathbf{p},\mathbf{q})$ is a Golay-ZCZ sequence pair of size $4N$ with $Z_{\min}=N$, if $x_1,x_2,x_3,x_4$ satisfy the following condition:
	\begin{equation}
		x_1x_2+x_3x_4=0.
	\end{equation}
\end{theorem}
\begin{IEEEproof} Let us consider $0\leq\tau\leq N-1$. Then we have
	\begin{equation}
		\begin{split}
			C_\mathbf{p}(\tau)=&2\big(C_\mathbf{a}(\tau) +C_\mathbf{b}(\tau)\big)\\&+(x_1x_2+x_3x_4)C_\mathbf{b,a}^*(N-\tau) \\&+x_2x_3C_\mathbf{a,b}^*(N-\tau), \\
			C_\mathbf{q}(\tau)=&2\big(C_\mathbf{c}(\tau) +C_\mathbf{d}(\tau)\big)\\&+(x_1x_2+x_3x_4)C_\mathbf{d,c}^*(N-\tau) \\&+x_2x_3C_\mathbf{c,d}^*(N-\tau).
		\end{split}
	\end{equation}
	Hence, for $0\leq\tau\leq N-1$, we have
	\begin{equation}
		C_\mathbf{p}(\tau)+C_\mathbf{q}(\tau)=4\big(C_\mathbf{a}(\tau) +C_\mathbf{b}(\tau)\big).
	\end{equation}
	
	Consider $N\leq\tau\leq 2N-1$. Then one has
	\begin{equation}
		\begin{split}
			C_\mathbf{p}(\tau)=& (x_1x_2+x_3x_4)C_\mathbf{a,b}(\tau-N) +x_2x_3C_\mathbf{b,a}(\tau-N) \\&+x_1x_3C_\mathbf{a}^*(2N-\tau) +x_2x_4C_\mathbf{b}^*(2N-\tau), \\
			C_\mathbf{q}(\tau)=& (x_1x_2+x_3x_4)C_\mathbf{c,d}(\tau-N) +x_2x_3C_\mathbf{d,c}(\tau-N) \\&+x_1x_3C_\mathbf{c}^*(2N-\tau) +x_2x_4C_\mathbf{d}^*(2N-\tau).
		\end{split}
	\end{equation}
	
	Hence, for $N\leq\tau\leq 2N-1$, we have
	\begin{equation}
		C_\mathbf{p}(\tau)+C_\mathbf{q}(\tau)=0.
	\end{equation}

	Consider $2N\leq\tau\leq 3N-1$. Then we have
	\begin{equation}
		\begin{split}
			C_\mathbf{p}(\tau)=& x_1x_3C_\mathbf{a}(\tau-2N) +x_2x_4C_\mathbf{b}(\tau-2N) \\&+x_1x_4C_\mathbf{b,a}^*(3N-\tau), \\
			C_\mathbf{q}(\tau)=& x_1x_3C_\mathbf{c}(\tau-2N) +x_2x_4C_\mathbf{d}(\tau-2N) \\&+x_1x_4C_\mathbf{d,c}^*(3N-\tau).
		\end{split}
	\end{equation}
	
	Hence, for $2N\leq\tau\leq 3N-1$, we have
	\begin{equation}
		C_\mathbf{p}(\tau)+C_\mathbf{q}(\tau)=0.
	\end{equation}
	
	Consider $3N\leq\tau\leq 4N-1$. Then we have
	\begin{equation}
		\begin{split}
			C_\mathbf{p}(\tau)=& x_1x_4C_\mathbf{a,b}(\tau-3N), \\
			C_\mathbf{q}(\tau)=& x_1x_4C_\mathbf{c,d}(\tau-3N).
		\end{split}
	\end{equation}
	
	Hence, for $3N\leq\tau\leq 4N-1$, we have
	\begin{equation}
		C_\mathbf{p}(\tau)+C_\mathbf{q}(\tau)=0.
	\end{equation}

	Hence, $(\mathbf{p,q})$ is a complementary pair of length $4N$.	
	
	Next, we have to prove the other two conditions of Definition \ref{golayzcz1d}. Consider $1\leq\tau\leq N$, then we have
	\begin{equation}
		\begin{split}
			R_\mathbf{p}(\tau)=& \sum_{k=0}^{4N-1} p_kp_{k+\tau}^*\\
			=& \sum_{k=0}^{N-1-\tau} a_ka_{k+\tau}^* +\sum_{k=N-\tau}^{N-1} x_1a_kx_2^*b_{k-(N-\tau)}^* \\&+\sum_{k=0}^{N-1-\tau} b_kb_{k+\tau}^* +\sum_{k=N-\tau}^{N-1} x_2b_kx_3^*a_{k-(N-\tau)}^* \\
			& +\sum_{k=0}^{N-1-\tau} a_ka_{k+\tau}^* +\sum_{k=N-\tau}^{N-1} x_3a_kx_4^*b_{k-(N-\tau)}^*\\& +\sum_{k=0}^{N-1-\tau} b_kb_{k+\tau}^* +\sum_{k=N-\tau}^{N-1} x_4b_kx_1^*a_{k-(N-\tau)}^* \\
			=&2\big(C_\mathbf{a}(\tau) +C_\mathbf{b}(\tau)\big). \\
		\end{split}
	\end{equation}
	Similarly for $1\leq\tau\leq N$, we have
	\begin{equation}
		\begin{split}
			R_\mathbf{q}(\tau)=&2\big(C_\mathbf{c}(\tau) +C_\mathbf{d}(\tau)\big), \\
			R_\mathbf{p,q}(\tau)=&2\big(C_\mathbf{a,c}(\tau) +C_\mathbf{b,d}(\tau)\big).
		\end{split}
	\end{equation}
	Since $(\mathbf{a},\mathbf{b})$ is a GCP and $(\mathbf{c},\mathbf{d})$ is one of the complementary mates of $(\mathbf{a},\mathbf{b})$, $R_\mathbf{p}(\tau)=R_\mathbf{q}(\tau)=0,\hbox{ for all } 1\leq\tau\leq N,$ and $R_\mathbf{p,q}(\tau)=0,\hbox{ for all } 0\leq\tau\leq N$.
	
	Therefore, $(\mathbf{p,q})$ is a $(2,4N,N)$- Golay-ZCZ sequence pair, consisting sequences of length $4N$ with $Z_{\min}=N$.
\end{IEEEproof}

\begin{example}\label{ex1 binary}
Let $(\mathbf{a,b})$ be a binary GCP of length 10, given by $\mathbf{a}=(1,1,-1,1,1,1,1,1,-1,-1)$, $\mathbf{b}=(1,1,-1,1,-1,1,-1, -1,1,1)$.
Then $(\mathbf{c,d})=(\overleftarrow{\mathbf{b}^*},-\overleftarrow{\mathbf{a}^*})$ is a Golay mate of $(\mathbf{a,b})$. Define
\begin{equation}
  \begin{split}
    \mathbf{p}=~&\mathbf{a}~\|~\mathbf{b}~\|~\mathbf{a}~\|-\mathbf{b}, \\
    \mathbf{q}=~&\mathbf{c}~\|~\mathbf{d}~\|~\mathbf{c}~\|-\mathbf{d}.
  \end{split}
\end{equation}
Then $(\mathbf{p,q})$ is $(2,40,10)$- Golay-ZCZ sequence pair, because
\begin{equation}
  \begin{split}
    \big(R_\mathbf{p}(\tau)\big)_{\tau=0}^{39}=&(40,\mathbf{0}_{10},-4,-8,4,8, -4,0,4,0,12,0,12,0,4,0,-4,8,4,-8,-4,\mathbf{0}_{10}), \\
    \big(R_\mathbf{q}(\tau)\big)_{\tau=0}^{39}=&(40,\mathbf{0}_{10},4,8,-4,-8, 4,0,-4,0,-12,0,-12,0,-4,0,4,-8,-4,8,4,\mathbf{0}_{10}), \\
    \big(R_\mathbf{p,q}(\tau)\big)_{\tau=0}^{39}=&(\mathbf{0}_{11},-4,-8,4, 16,4,0,4,-8,-4,0,4,-8,12,0,12,0,-4,8,4,\mathbf{0}_{10}).
  \end{split}
\end{equation}
\end{example}

\begin{example}\label{ex1}
Let $\mathbf{a}=(1,i,-i,-1,i),\mathbf{b}=(1,1,1,i,-i)$, then $(\mathbf{a,b})$ be a quadriphase GCP of length 5.
Then $(\mathbf{c,d})=(\overleftarrow{\mathbf{b}^*},-\overleftarrow{\mathbf{a}^*})$ is a Golay mate of $(\mathbf{a,b})$. Define
\begin{equation}
  \begin{split}
    \mathbf{p}=~&\mathbf{a}~\|~\mathbf{b}~\|~\mathbf{a}~\|-\mathbf{b},\\
    \mathbf{q}=~&\mathbf{c}~\|~\mathbf{d}~\|~\mathbf{c}~\|-\mathbf{d}.
  \end{split}
\end{equation}
Then $(\mathbf{p,q})$ is a $(2,20,5)$- Golay-ZCZ sequence pair. The periodic correlation magnitudes of $(2,20,5)$- Golay-ZCZ sequence pair ($\mathbf{p,q}$) are shown in Fig. \ref{fig1}.
\begin{figure}[ht]
  \centering
  \includegraphics[width=.7\textwidth]{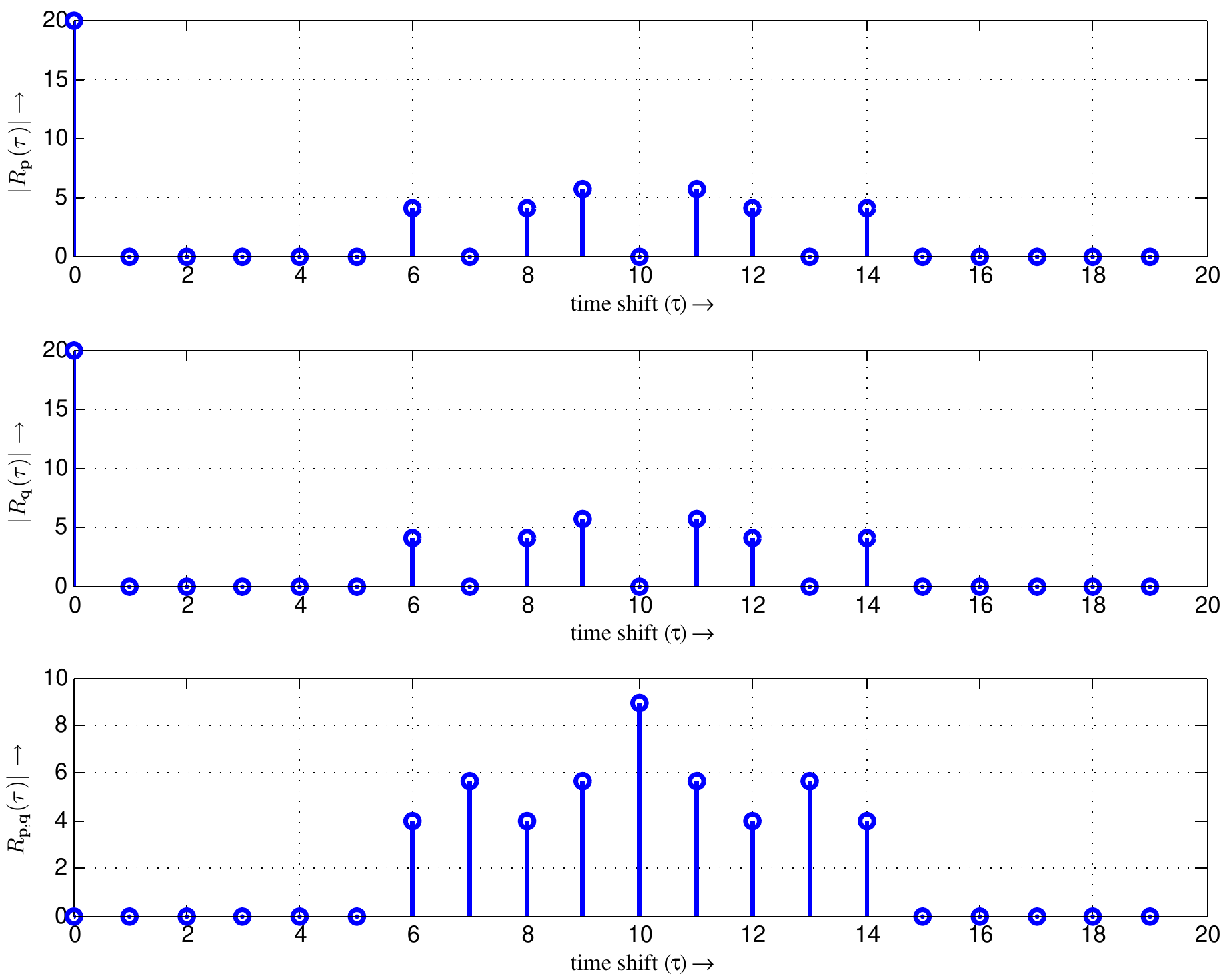}
  \caption{A glimpse of the periodic correlations of the proposed Golay-ZCZ sequence pair given in Example \ref{ex1}.}\label{fig1}
\end{figure}
\end{example}

\begin{remark}
	Please note that binary Golay-ZCZ sequences of length 40 and quadriphase Golay-ZCZ sequences of length 20 has not been previously reported in the literature. By considering a GCP $(\mathbf{a,b})$ of length $2^m$ in Theorem \ref{th1}, we can construct $(\mathbf{p,q})$ of length $2^{m+2}$ and the resultant Golay-ZCZ sequences will have the parameters equivalent to the Golay-ZCZ sequences reported in \cite{Chen201811}.
\end{remark}

\section{New Construction of Complementary Sets with ZCZ}\label{golayzcz2d}
The resultant sequence sets of the constructions reported in \cite{Chen201811} have the property that, as the number of sequences of the Golay-ZCZ sequence set increases, the ZCZ width decreases. Also, the resultant sequence sets in \cite{Chen201811} are not optimal for polyphase cases. These problems are taken care of in the following construction. In this section, we propose a new construction of Golay-ZCZ sequence sets based on CCCs, which are asymptotically optimal. Before we proceed further, we reveal a nice property of CCC.

\begin{property}\label{prop1}
	Let $\mathfrak{C}=\left\{\mathcal{C}^{0}, \mathcal{C}^{1}, \cdots, \mathcal{C}^{M-1}\right\}$ be a $(M,M,N)$- CCC, consisting $M$ CSs $\mathcal{C}^{k}$, $0\leq k<M$, each having $M$ sequences of length $N$. Let $\mathcal{D}$ be a sequence set of order $M\times MN$ defined as follows
	\begin{equation}
		\mathcal{D}=\begin{bmatrix}
			\mathbf{c}^0_0 & \mathbf{c}^0_1 & \cdots & \mathbf{c}^0_{M-1}\\
			\mathbf{c}^1_0 & \mathbf{c}^1_1 & \cdots & \mathbf{c}^1_{M-1}\\
			\vdots & \vdots & \ddots & \vdots\\
			\mathbf{c}^{M-1}_0 & \mathbf{c}^{M-1}_1 & \cdots & \mathbf{c}^{M-1}_{M-1}
		\end{bmatrix}_{M\times MN},
	\end{equation}
where the $i$-th row is generated from $\mathcal{C}^i$, by appending the rows of $\mathcal{C}^i$ to the right of one another. for $0\leq i \leq M-1$, let
\begin{equation}
	\mathcal{D}^i=\begin{bmatrix}
		\mathbf{c}^0_i\\ \mathbf{c}^1_i\\ \vdots\\ \mathbf{c}^{M-1}_i
	\end{bmatrix}.
\end{equation}Then $\mathfrak{D}=\{\mathcal{D}^0,\mathcal{D}^1,\cdots,\mathcal{D}^{M-1}\}$ is also an $(M,M,N)$- CCC.
\end{property}

\begin{IEEEproof}
	$\mathbf{c}_i^{k}(x)$ denote characteristic polynomial of sequence $\mathbf{c}_i^{k}$. As $\mathcal{C}^{k}=\{\mathbf{c}_0^{k},\mathbf{c}_1^{k},\cdots,\mathbf{c}_{M-1}^{k}\}$ is a CS, then we have
	\begin{equation}
		\sum_{i=0}^{M-1}\mathbf{c}_i^{k}(x)\left(\mathbf{c}_i^{k}(x^{-1})\right)^*=MN, \hbox{ for any }m.
	\end{equation}
	As $\mathfrak{C}=\left\{\mathcal{C}^{0}, \mathcal{C}^{1}, \cdots, \mathcal{C}^{M-1}\right\}$ is a $(M,M,N)$- CCC, we have
	\begin{equation}
		\sum_{i=0}^{M-1}\mathbf{c}_i^{k_1}(x)\left(\mathbf{c}_i^{k_2}(x^{-1})\right)^*=0, \hbox{ if }k_1 \neq k_2.
	\end{equation}
	By matrix representation, we have
\begin{equation}
	\left(
	\begin{array}{ccc}
		\mathbf{c}_0^{0}(x) & \cdots & \mathbf{c}_{M-1}^{0}(x) \\
		\vdots & \ddots & \vdots \\
		\mathbf{c}_0^{M-1}(x) & \cdots & \mathbf{c}_{M-1}^{M-1}(x) \\
	\end{array}
	\right)
	\left(
	\begin{array}{ccc}
		\mathbf{c}_0^{0}(x^{-1}) & \cdots & \mathbf{c}_{0}^{M-1}(x^{-1}) \\
		\vdots & \ddots & \vdots \\
		\mathbf{c}_{M-1}^{0}(x^{-1}) & \cdots & \mathbf{c}_{M-1}^{M-1}(x^{-1}) \\
	\end{array}
	\right)^*
	=
	\left(
	\begin{array}{ccc}
		MN & \cdots & 0 \\
		\vdots & \ddots & \vdots \\
		0 & \cdots & MN \\
	\end{array}
	\right).
\end{equation}
	By the commutative law of matrix, we have
	\begin{equation}
		\left(
		\begin{array}{ccc}
			\mathbf{c}_0^{0}(x^{-1}) & \cdots & \mathbf{c}_{0}^{M-1}(x^{-1}) \\
			\vdots & \ddots & \vdots \\
			\mathbf{c}_{M-1}^{0}(x^{-1}) & \cdots & \mathbf{c}_{M-1}^{M-1}(x^{-1}) \\
		\end{array}
		\right)^*
		\left(
		\begin{array}{ccc}
			\mathbf{c}_0^{0}(x) & \cdots & \mathbf{c}_{M-1}^{0}(x) \\
			\vdots & \ddots & \vdots \\
			\mathbf{c}_0^{M-1}(x) & \cdots & \mathbf{c}_{M-1}^{M-1}(x) \\
		\end{array}
		\right)
		=
		\left(
		\begin{array}{ccc}
			MN & \cdots & 0 \\
			\vdots & \ddots & \vdots \\
			0 & \cdots & MN \\
		\end{array}
		\right).
	\end{equation}
	Let $y=x^{-1}$, we have
	\begin{equation}
		\left(
		\begin{array}{ccc}
			\mathbf{c}_0^{0}(y) & \cdots & \mathbf{c}_{0}^{M-1}(y) \\
			\vdots & \ddots & \vdots \\
			\mathbf{c}_{M-1}^{0}(y) & \cdots & \mathbf{c}_{M-1}^{M-1}(y) \\
		\end{array}
		\right)^*
		\left(
		\begin{array}{ccc}
			\mathbf{c}_0^{0}(y^{-1}) & \cdots & \mathbf{c}_{M-1}^{0}(y^{-1}) \\
			\vdots & \ddots & \vdots \\
			\mathbf{c}_0^{M-1}(y^{-1}) & \cdots & \mathbf{c}_{M-1}^{M-1}(y^{-1}) \\
		\end{array}
		\right)
		=
		\left(
		\begin{array}{ccc}
			MN & \cdots & 0 \\
			\vdots & \ddots & \vdots \\
			0 & \cdots & MN \\
		\end{array}
		\right).
	\end{equation}
	Taking conjugation on the above equation, we have
	\begin{equation}
		\left(
		\begin{array}{ccc}
			\mathbf{c}_0^{0}(y) & \cdots & \mathbf{c}_{0}^{M-1}(y) \\
			\vdots & \ddots & \vdots \\
			\mathbf{c}_{M-1}^{0}(y) & \cdots & \mathbf{c}_{M-1}^{M-1}(y) \\
		\end{array}
		\right)
		\left(
		\begin{array}{ccc}
			\mathbf{c}_0^{0}(y^{-1}) & \cdots & \mathbf{c}_{M-1}^{0}(y^{-1}) \\
			\vdots & \ddots & \vdots \\
			\mathbf{c}_0^{M-1}(y^{-1}) & \cdots & \mathbf{c}_{M-1}^{M-1}(y^{-1}) \\
		\end{array}
		\right)^*
		=
		\left(
		\begin{array}{ccc}
			MN & \cdots & 0 \\
			\vdots & \ddots & \vdots \\
			0 & \cdots & MN \\
		\end{array}
		\right).
	\end{equation}
	Hence proved.
\end{IEEEproof}

\begin{theorem}\label{const2n}
	Let $\mathfrak{C}=\{\mathcal{C}^0,\mathcal{C}^1,\dots,\mathcal{C}^{M-1}\}$ be an $(M,M,N)$- CCC, and $\mathcal{F}$ be an IDFT matrix of order $M\times M$, where $f_{i,j}$ denotes the element of $i$-th row and $j$-th column of $\mathcal{F}$. Define $\mathcal{B}_k$ for $0\leq k<M$, as follows
	\begin{equation}
		\mathcal{B}_k=\begin{bmatrix}
			f_{0,0}\mathbf{c}_{k}^0 & f_{0,1}\mathbf{c}_{k}^1 & \cdots & f_{0,M-1}\mathbf{c}_{k}^{M-1}\\
			f_{1,0}\mathbf{c}_{k}^0 & f_{1,1}\mathbf{c}_{k}^1 & \cdots & f_{1,M-1}\mathbf{c}_{k}^{M-1}\\
			\vdots & \vdots & \ddots &  \vdots \\
			f_{M-1,0}\mathbf{c}_{k}^0 & f_{M-1,1}\mathbf{c}_{k}^1 & \cdots & f_{M-1,M-1}\mathbf{c}_{k}^{M-1}\\
		\end{bmatrix}_{M\times MN}
	\end{equation}
Let $\tilde{\mathcal{B}_k}$ denotes a sequence of length $M^2N$ which is generated from $\mathcal{B}_k$, by appending the rows of $\mathcal{B}_k$ to the right of one another. Then the sequence set $\mathcal{D}=\{\tilde{\mathcal{B}}_0,\tilde{\mathcal{B}}_1,\dots,\tilde{\mathcal{B}}_{M-1}\}$ is a CS of length $M^2N$ with $Z_{\min}=(M-1)N$. 
\end{theorem}


\begin{IEEEproof}
	First we prove that $\mathcal{D}$ is a CS. For $0 < \tau \leq N-1$, we have
	\begin{equation}
		\begin{split}
			\sum_{i=0}^{M-1}C_{\tilde{\mathcal{B}}_i}(\tau)=&\sum_{j=0}^{M-1}\left[\left(\sum_{k=0}^{M-1}f^2_{k,j}\right)\left(\sum_{k=0}^{M-1}C_{\mathbf{c}^j_k}(\tau)\right)\right]+\sum_{j=0}^{M-2}\left[\left(\sum_{k=0}^{M-1}f_{k,j}f^*_{k,j+1}\right)\left(\sum_{k=0}^{M-1}C_{\mathbf{c}^j_k,\mathbf{c}^{j+1}_k}(\tau)\right)\right]\\&+\left(\sum_{k=0}^{M-2}f_{k,M-1}f^*_{k+1,0}\right)\left(\sum_{k=0}^{M-1}C_{\mathbf{c}^{M-1}_k,\mathbf{c}^0_k}(\tau)\right).
		\end{split}
	\end{equation}
Since $\mathfrak{C}=\{\mathcal{C}^0,\mathcal{C}^1,\dots,\mathcal{C}^{M-1}\}$ is a CCC, we have $\sum_{i=0}^{M-1}C_{\tilde{\mathcal{B}}_i}(\tau)=0$. Similarly, we can show for other values of $\tau$ that $\sum_{i=0}^{M-1}C_{\tilde{\mathcal{B}}_i}(\tau)=0$.

Next, we prove the other two conditions of Definition \ref{golayzcz1d}. For $0\leq \tau < M^2N$, let $\tau=rN+s$. Let us define
\begin{equation}
	\mathcal{H}_{\mathcal{B}_{k_1},\mathcal{B}_{k_2}}={\mathcal{B}_{k_1}} \odot L^r_s({\mathcal{B}_{k_2}}),
\end{equation}
where $\odot$ denotes elementwise product of the matrices $\mathcal{B}_{k_1}$ and $L^r_s({\mathcal{B}_{k_2}})$. Note that $\mathcal{H}_{\mathcal{B}_{k_1},\mathcal{B}_{k_2}}$ is a matrix of size $M\times N$. When $k_1=k_2$, we write $\mathcal{H}_{\mathcal{B}_{k_1}}$ instead of $\mathcal{H}_{\mathcal{B}_{k_1},\mathcal{B}_{k_2}}$. $\sum \mathcal{H}_{\mathcal{B}_{k_1},\mathcal{B}_{k_2}}$ denotes sum of all elements of $\mathcal{H}_{\mathcal{B}_{k_1},\mathcal{B}_{k_2}}$. To check the periodic autocorrelation of the constituent sequences of $\mathcal{D}$, we have, for $0<\tau \leq (M-1)N+1$ and $0\leq k <M$,
\begin{equation}
	\begin{split}
		R_{\tilde{\mathcal{B}_k}}(\tau)&=\sum \mathcal{H}_{\mathcal{B}_{k}}\\
		&=\sum_{i=0}^{M-1} (f_{:,i})\cdot (T^{\lfloor\frac{i+r}{M}\rfloor}(f_{:,<i+r>_M})) C_{\mathbf{c}_k^{i},\mathbf{c}_k^{<i+r>_M}}(\tau-rN) \\
		&+\sum_{i=0}^{M-1} (f_{:,i})\cdot (T^{\lfloor\frac{i+r+1}{M}\rfloor}(f_{:,<i+r+1>_M})) C_{\mathbf{c}_k^{i},\mathbf{c}_k^{<i+r+1>_M}}(\tau-(r+1)N), 
	\end{split}
\end{equation} 
where `$\cdot$' denotes the `inner product' of two sequences, and $<x>_M$ denotes $x \mod M$. Since $\mathcal{F}$ is an IDFT matrix, using property \ref{prop1}, we have
\begin{equation}
	R_{\tilde{\mathcal{B}_k}}(\tau)=\begin{cases}
		M^2N, & \text{when }\tau=0;\\
		0, & \text{when }1\leq \tau \leq (M-1)N;\\
		\text{non-zero} & \text{when }(M-1)N< \tau.
	\end{cases}
\end{equation} 
Similarly, to check the cross-correlation we have for $0\leq k_1\neq k_2 <M$,

\begin{equation}
	\begin{split}
		R_{\tilde{\mathcal{B}}_{k_1},\tilde{\mathcal{B}}_{k_2}}(\tau)&=\sum \mathcal{H}_{\mathcal{B}_{k_1},\mathcal{B}_{k_2}}\\
		&=\sum_{i=0}^{M-1} (f_{:,i})\cdot (T^{\lfloor\frac{i+r}{M}\rfloor}(f_{:,<i+r>_M})) C_{\mathbf{c}_{k_1}^{i},\mathbf{c}_{k_2}^{<i+r>_M}}(\tau-rN) \\
		&+\sum_{i=0}^{M-1} (f_{:,i})\cdot (T^{\lfloor\frac{i+r+1}{M}\rfloor}(f_{:,<i+r+1>_M})) C_{\mathbf{c}_{k_1}^{i},\mathbf{c}_{k_2}^{<i+r+1>_M}}(\tau-(r+1)N).
	\end{split}
\end{equation} 
Since $\mathfrak{C}$ is a CCC and $\mathcal{F}$ is an IDFT matrix, using property \ref{prop1}, we have
 \begin{equation}
 	R_{\tilde{\mathcal{B}}_{k_1},\tilde{\mathcal{B}}_{k_2}}(\tau)=\begin{cases}
 		0, & \text{when }\tau=0;\\
 		0, & \text{when }1\leq \tau \leq (M-1)N;\\
 		\text{non-zero} & \text{when }(M-1)N< \tau.
 	\end{cases}
 \end{equation} 
Hence, $\mathcal{D}$ is a $(M,M^2N,(M-1)N)$- Golay-ZCZ sequence set of set size $M$, consisting sequences of length $M^2N$ having ZCZ width $(M-1)N$.
\end{IEEEproof}
\begin{example}\label{ex3n}
	Let $\mathfrak{C}=\{\mathcal{C}^0,\mathcal{C}^1,\mathcal{C}^2,\mathcal{C}^3\}$ be a $(4,4,4)$- CCC given by
	\begin{equation}
		\begin{split}
			\mathcal{C}^0=\begin{bmatrix}
				1 & 1 & 1 & 1\\
				1 & 1 & -1 & -1\\
				-1 & 1 & -1 & 1\\
				-1 & 1 & 1 & -1
			\end{bmatrix},&
			\mathcal{C}^1=\begin{bmatrix}
				-1 & -1 & 1 & 1\\
				-1 & -1 & -1 & -1\\
				1 & -1 & -1 & 1\\
				1 & -1 & 1 & -1
			\end{bmatrix},\\
			\mathcal{C}^2=\begin{bmatrix}
				-1 & 1 & -1 & 1\\
				-1 & 1 & 1 & -1\\
				1 & 1 & 1 & 1\\
				1 & 1 & -1 & -1
			\end{bmatrix},&
			\mathcal{C}^3=\begin{bmatrix}
				1 & -1 & -1 & 1\\
				1 & -1 & 1 & -1\\
				-1 & -1 & 1 & 1\\
				-1 & -1 & -1 & -1
			\end{bmatrix},
		\end{split}
	\end{equation}
	and $\mathcal{F}$ be an $4\times 4$ IDFT matrix. Construct $\mathcal{D}=\{\tilde{\mathcal{B}}_0,\tilde{\mathcal{B}}_1,\tilde{\mathcal{B}}_2,\tilde{\mathcal{B}}_3\}$ as per Construction \ref{const2n}. Then $\mathcal{D}$ is a $(4,64,12)$- Golay-ZCZ sequence set. A glimpse of the correlations of the sequence set $\mathcal{D}$ is shown in Fig. \ref{fig2}.
	\begin{figure}[ht]
		\centering
		\includegraphics[width=.7\textwidth]{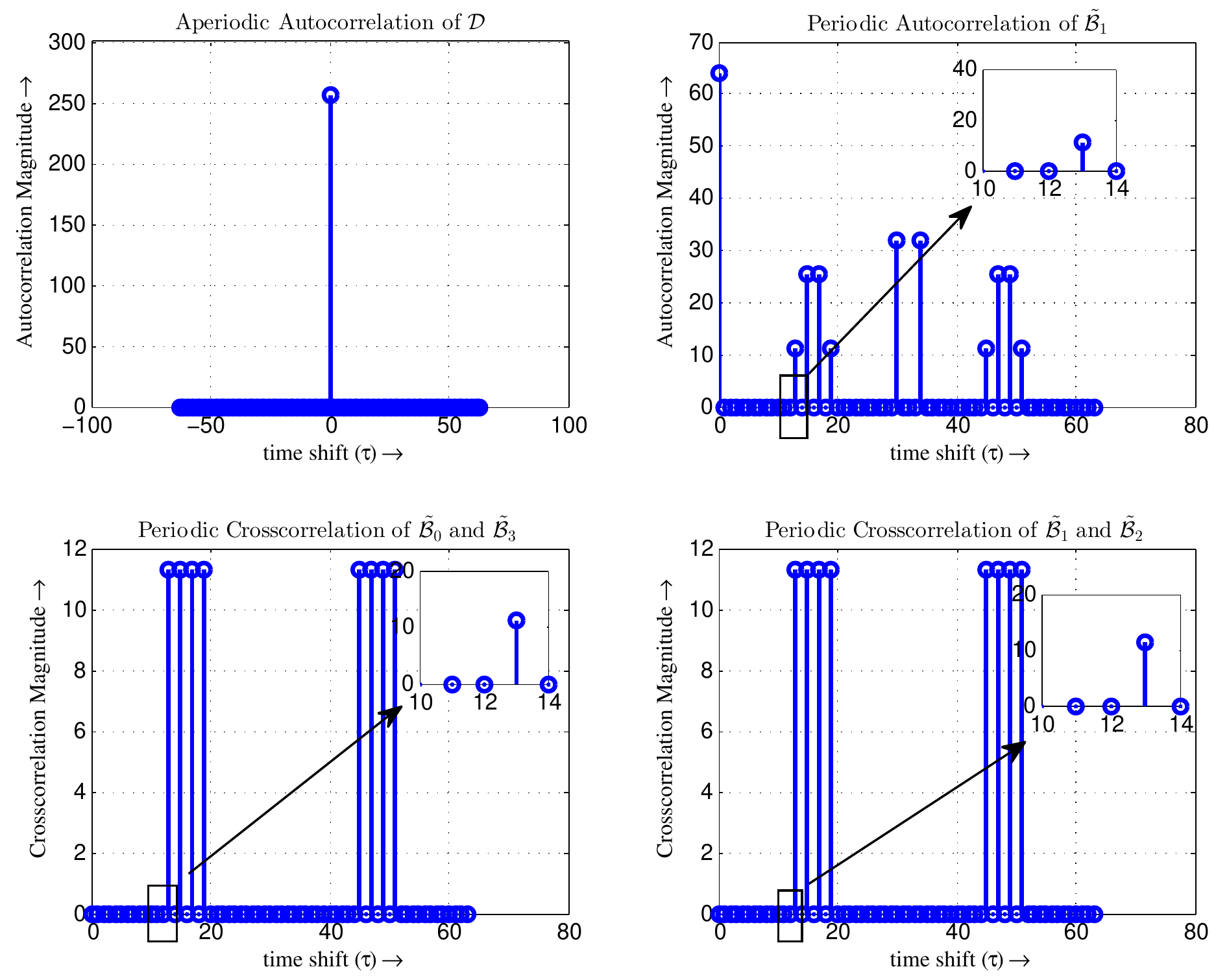}
		\caption{A glimpse of the correlations of the sequence set $\mathcal{D}$, constructed in Example \ref{ex3n}.}\label{fig2}
	\end{figure}
\end{example}
\begin{remark}
	In \cite{Chen201811}, the authors reported $(2^k,2^m,2^{m-k-1})$- Golay-ZCZ sequence sets. Considering $k=2$, and $m=6$, we get $(4,64,8)$- Golay-ZCZ sequence set. As we see in Example \ref{ex3n}, the Golay-ZCZ sequence sets proposed in Theorem \ref{const2n} have larger ZCZ widths, as compared to the results in \cite{Chen201811}. 
\end{remark}


Since Theorem \ref{const2n} is based on CCCs, the availability of CCCs for various flexible lengths highly improves the outcome of the construction. In Table \ref{tabccc}, we list down all the well-known construction of CCCs till date. We also provide an iterative construction of CCC to improve the flexibility of the parameters. Before we proceed, we need the following lemma.

\begin{lemma}\label{lemnew}
	Let $(\mathbf{x}_1,\mathbf{x}_2)$ be two sequences of length $N_1$, $(\mathbf{y}_1,\mathbf{y}_2)$ be two sequences of length $N_2$, then
	aperiodic correlation of $\mathbf{x}_1\otimes\mathbf{y}_1$ between $\mathbf{x}_2\otimes\mathbf{y}_2$ is given by \cite{jin}
	\begin{equation}
		C_{\mathbf{x}_1\otimes\mathbf{y}_1,\mathbf{x}_2\otimes\mathbf{y}_2} (\tau)= C_{\mathbf{x}_1,\mathbf{x}_2}(k_1) C_{\mathbf{y}_1,\mathbf{y}_2}(k_2) +C_{\mathbf{x}_1,\mathbf{x}_2}(k_1+1)C_{\mathbf{y}_1,\mathbf{y}_2}(k_2-N),
	\end{equation}
	where $\tau=k_1N_2+k_2$.
\end{lemma}

\begin{theorem}\label{th3}
	Let $\mathfrak{C}=\{\mathcal{C}^0,\mathcal{C}^1,\dots,\mathcal{C}^{M-1}\}$ and $\mathfrak{D}=\{\mathcal{D}^0,\mathcal{D}^1,\dots,\mathcal{D}^{M-1}\}$ be two CCCs with parameters $(M,M,N_1)$ and $(M,M,N_2)$, respectively. Then $\mathfrak{E}=\{\mathcal{E}^0,\mathcal{E}^1,\dots,\mathcal{E}^{M-1}\}$, given by
	\begin{equation}
		\mathcal{E}^{m}=\begin{bmatrix}
			\mathbf{e}^{m}_0\\
			\mathbf{e}^{m}_1\\
			\vdots\\
			\mathbf{e}^{m}_{M-1}
		\end{bmatrix}_{M \times MN_1N_2}=\left[
	\begin{array}{cccc}
	\mathbf{c}^{m}_0\otimes\mathbf{d}^{0}_0 & \mathbf{c}^{m}_1\otimes\mathbf{d}^{0}_1 & \cdots & \mathbf{c}^{m}_{M-1}\otimes\mathbf{d}^{0}_{M-1} \\
	\mathbf{c}^{m}_0\otimes\mathbf{d}^{1}_0 & \mathbf{c}^{m}_1\otimes\mathbf{d}^{1}_1 & \cdots & \mathbf{c}^{m}_{M-1}\otimes\mathbf{d}^{1}_{M-1} \\
	\vdots & \vdots &  & \vdots \\
	\mathbf{c}^{m}_0\otimes\mathbf{d}^{M-1}_0 & \mathbf{c}^{m}_1\otimes\mathbf{d}^{M-1}_1 & \cdots & \mathbf{c}^{m}_{M-1}\otimes\mathbf{d}^{M-1}_{M-1} \\
\end{array}
\right].
	\end{equation}	
is a CCC with parameters $(M,M,MN_1,N_2)$.	
\end{theorem}

\begin{IEEEproof}

First, we prove that for $0\leq m <M$, $\mathcal{E}^m$ is a CS of length $MN_1N_2$.

By lemma \ref{lemnew}, we have
\begin{equation}
	\begin{split}
		\sum_{l=0}^{M-1}C_{\mathbf{c}^{m}_{n_1}\otimes\mathbf{d}^{l}_{n_1}, \mathbf{c}^{m}_{n_2}\otimes\mathbf{d}^{l}_{n_2}} (\tau_0) = & \sum_{l=0}^{M-1} \left(C_{\mathbf{c}^{m}_{n_1}, \mathbf{c}^{m}_{n_2}}(k_1) C_{\mathbf{d}^{l}_{n_1}, \mathbf{d}^{l}_{n_2}}(k_2)+C_{\mathbf{c}^{m}_{n_1}, \mathbf{c}^{m}_{n_2}}(k_1+1) C_{\mathbf{d}^{l}_{n_1}, \mathbf{d}^{l}_{n_2}}(k_2-N_2)\right) \\
		= & C_{\mathbf{c}^{m}_{n_1}, \mathbf{c}^{m}_{n_2}}(k_1)\sum_{l=0}^{M-1} C_{\mathbf{d}^{l}_{n_1}, \mathbf{d}^{l}_{n_2}}(k_2)+C_{\mathbf{c}^{m}_{n_1}, \mathbf{c}^{m}_{n_2}}(k_1+1) \sum_{l=0}^{M-1} C_{\mathbf{d}^{l}_{n_1}, \mathbf{d}^{l}_{n_2}}(k_2-N_2)\\
		= & \begin{cases}
			MN_2C_{\mathbf{c}^{m}_{n_1}, \mathbf{c}^{m}_{n_2}}(k_1), & k_2=0,n_1=n_2; \\
			0, & k_2\neq0,n_1=n_2; \\
			0, & \hbox{for all }k_2,n_1\neq n_2,
		\end{cases}
	\end{split}
\end{equation}
where $\tau_0=k_1N_2+k_2$.
For $\mathcal{E}^m$, let us consider $\tau=rN_1N_2+\tau_0$, then
\begin{equation}\label{eq39n}
	\sum_{l=0}^{M-1}C_{\mathbf{e}^m_l}(\tau)= \sum_{l=0}^{M-1}\sum_{n_1=0}^{M-1-r}(C_{\mathbf{c}^{m}_{n_1}\otimes\mathbf{d}^{l}_{n_1}, \mathbf{c}^{m}_{n_2}\otimes\mathbf{d}^{l}_{n_2}} (\tau_0)+ C_{\mathbf{c}^{m}_{n_1}\otimes\mathbf{d}^{l}_{n_1}, \mathbf{c}^{m}_{n_2+1}\otimes\mathbf{d}^{l}_{n_2+1}} (\tau_0-N_1N_2)),
\end{equation}
where $n_2=n_1+r$.

Clearly, we have $\sum_{l=0}^{M-1}C_{\mathbf{e}^m_l}(\tau)=0$, if $r\geq1$. Consider $r=0$, then
\begin{equation}
	\begin{split}
		\sum_{l=0}^{M-1}C_{\mathbf{e}^m_l}(\tau) = & \sum_{l=0}^{M-1}C_{\mathbf{e}^m_l}(\tau_0) \\
		= & \sum_{l=0}^{M-1}\sum_{n_1=n_2=0}^{M-1} C_{\mathbf{c}^{m}_{n_1}\otimes\mathbf{d}^{l}_{n_1}, \mathbf{c}^{m}_{n_2}\otimes\mathbf{d}^{l}_{n_2}} (\tau_0) \\
		= & \left\{
		\begin{array}{ll}
			\sum\limits_{n_1=n_2=0}^{M-1}MN_2C_{\mathbf{c}^{m}_{n_1}, \mathbf{c}^{m}_{n_2}}(k_1), & k_2=0; \\
			0, & k_2\neq 0.
		\end{array}
		\right. \\
		= & \left\{
		\begin{array}{ll}
			M^2N_1N_2,  & k_1=0,k_2=0; \\
			0, & \hbox{otherwise.}
		\end{array}
		\right.
	\end{split}
\end{equation}

Next, we prove that $\mathcal{E}^{m_1}$ and $\mathcal{E}^{m_2}$ are orthogonal if $m_1\neq m_2$.

Similar to (\ref{eq39n}), we consider cross-correlation sum of $\mathcal{E}^{m_1}$ between $\mathcal{E}^{m_2}$, then we have

\begin{equation}
	\sum_{l=0}^{M-1}C_{\mathbf{e}^{m_1}_l,\mathbf{e}^{m_2}_l}(\tau)= \sum_{l=0}^{M-1}\sum_{n_1=0}^{M-1-r}(C_{\mathbf{c}^{m_1}_{n_1}\otimes\mathbf{d}^{l}_{n_1}, \mathbf{c}^{m_2}_{n_2}\otimes\mathbf{d}^{l}_{n_2}} (\tau_0)+ C_{\mathbf{c}^{m_1}_{n_1}\otimes\mathbf{d}^{l}_{n_1}, \mathbf{c}^{m_2}_{n_2+1}\otimes\mathbf{d}^{l}_{n_2+1}} (\tau_0-N_1N_2)),
\end{equation}
where $n_2=n_1+r$.

Clearly, we have $\sum_{l=0}^{M-1}C_{\mathbf{e}^m_l}(\tau)=0$, if $r\geq1$. Consider $r=0$, then
\begin{equation}
	\begin{split}
		\sum_{l=0}^{M-1}C_{\mathbf{e}^{m_1}_l,\mathbf{e}^{m_2}_l}(\tau) = & \sum_{l=0}^{M-1}C_{\mathbf{e}^{m_1}_l,\mathbf{e}^{m_2}_l}(\tau_0) \\
		= & \sum_{l=0}^{M-1}\sum_{n_1=n_2=0}^{M-1} C_{\mathbf{c}^{m_1}_{n_1}\otimes\mathbf{d}^{l}_{n_1}, \mathbf{c}^{m_2}_{n_2}\otimes\mathbf{d}^{l}_{n_2}} (\tau_0) \\
		= & \left\{
		\begin{array}{ll}
			\sum\limits_{n_1=n_2=0}^{M-1}MN_2C_{\mathbf{c}^{m_1}_{n_1}, \mathbf{c}^{m_2}_{n_2}}(k_1), & k_2=0; \\
			0, & k_2\neq 0.
		\end{array}
		\right. \\
		= & 0.
	\end{split}
\end{equation}
Hence $\mathfrak{E}=\{\mathcal{E}^0,\mathcal{E}^1,\dots,\mathcal{E}^{M-1}\}$ is an $(M,M,MN_1N_2)$- CCC.
\end{IEEEproof}

\begin{remark}\label{remarkccc}
	To increase the flexibility of the proposed construction in Theorem \ref{const2n}, we have also found binary CCCs with parameters $(4,4,3)$, $(4,4,5)$, $(4,4,7)$, $(4,4,11)$, and $(4,4,13)$ based on computer search, which can be used as seed CCCs of the proposed construction in Theorem \ref{const2n}. Considering $\mathfrak{C}=\{\mathcal{C}^0,\mathcal{C}^1,\mathcal{C}^2,\mathcal{C}^3\}$ as a $(4,4,N)$- CCC, the search results can be found in Table \ref{tab3}, where each element represents a power of $(-1)$. The search results are important in itself, because in recent results \cite{chenccc,yuboccc,bingshenccc}, we observe that for a $(K,M,N)$ mutually orthogonal sequence set, through systematic construction, the maximum achievable $K/M$ ratio is $1/2$, when $N$ is not in the form of $2^m$. However, for our case, although $N$ is not in the form of power-of-two, since the sequence sets are CCC (i.e., $K=M$), the $K/M$ ratio is $1$. Moreover, for length up to 200 (i.e., $N\leq 200$), using the CCCs given in Table \ref{tab3} as seed CCCs, using the results of \cite{jin} and Theorem \ref{th3}, we can design binary $(4,4,N)$- CCCs for $N=12,~13,~20,~24,~28,~36,~40,~44,~48,~52,~56,~60$, $72,~80,~84,~88,~96,~112,~120,~132,~140$, $144,~156,~160,~168,~176,~192,~196,~200$.
	\end{remark}

\begin{table}[]
		\renewcommand{\arraystretch}{1.3}
	\resizebox{\textwidth}{!}{
		\begin{tabular}{|c|c|c|c|c|}
			\hline
			$N$                                                                                                                                                   & $\mathcal{C}^{0}$                                                                                                                                                   & $\mathcal{C}^{1}$                                                                                                                                                   & $\mathcal{C}^{2}$                                                                                                                                                   & $\mathcal{C}^{3}$                                                                                                                                                                                                                                                                                                      \\ \hline
			3 & \begin{tabular}[c]{@{}c@{}}000\\
				001\\
				001\\
				010\end{tabular} & \begin{tabular}[c]{@{}c@{}}010\\
				001\\
				110\\
				111\end{tabular} & \begin{tabular}[c]{@{}c@{}}011\\
				000\\
				101\\
				100\end{tabular} & \begin{tabular}[c]{@{}c@{}}011\\
				010\\
				000\\
				011\end{tabular}  \\ \hline
			5 & \begin{tabular}[c]{@{}c@{}}00001\\
				01100\\
				01000\\
				01011\end{tabular} & \begin{tabular}[c]{@{}c@{}}00010\\
				00101\\
				01111\\
				00110\end{tabular} & \begin{tabular}[c]{@{}c@{}}00101\\
				11101\\
				00110\\
				10000\end{tabular} & \begin{tabular}[c]{@{}c@{}}01100\\
				11110\\
				01011\\
				10111\end{tabular}  \\ \hline
			7 & \begin{tabular}[c]{@{}c@{}}0000001\\
				0011010\\
				0011010\\
				0100011\end{tabular} & \begin{tabular}[c]{@{}c@{}}0011101\\
				0011010\\
				1100101\\
				1000000\end{tabular} & \begin{tabular}[c]{@{}c@{}}0101100\\
				0100011\\
				1111110\\
				1010011\end{tabular} & \begin{tabular}[c]{@{}c@{}}0101100\\
				0111111\\
				0011101\\
				0101100\end{tabular}  \\ \hline
			11 & \begin{tabular}[c]{@{}c@{}}01110110110\\
				00111000101\\
				00011010100\\
				00000001101\end{tabular} & \begin{tabular}[c]{@{}c@{}}00011010100\\
				00000001101\\
				10001001001\\
				11000111010\end{tabular} & \begin{tabular}[c]{@{}c@{}}10110000000\\
				11010100111\\
				10100011100\\
				10010010001\end{tabular} & \begin{tabular}[c]{@{}c@{}}01011100011\\
				01101101110\\
				10110000000\\
				11010100111\end{tabular}  \\ \hline
			13 & \begin{tabular}[c]{@{}c@{}}0111011010100\\
				0011101001101\\
				0001100001001\\
				0000000111010\end{tabular} & \begin{tabular}[c]{@{}c@{}}0001100001001\\
				0000000111010\\
				1000100101011\\
				1100010110010\end{tabular} & \begin{tabular}[c]{@{}c@{}}0101110000000\\
				0110111100111\\
				1011001011100\\
				1101010010001\end{tabular} & \begin{tabular}[c]{@{}c@{}}0100110100011\\
				0010101101110\\
				0101110000000\\
				0110111100111\end{tabular}  \\ \hline
	\end{tabular}}
	\caption{Computer search results of $(4,4,N)$- CCCs, for various values of $N$.\label{tab3}}
\end{table}

\section{Discussion on Optimality of the Proposed Sequence Sets}\label{secv}
 For polyphase sequence sets consisting of $M$ sequences each of length $L$, having ZCZ width $Z$, we have from \cite{tangfan}
\begin{equation}
	Z\leq \frac{L}{M}.
\end{equation}
For binary sequence sets, the bound is conjectured \cite{tangfan} as 
\begin{equation}
	Z\leq \frac{L}{2M}.
\end{equation}
Assuming $Z_{opti}$ to be the optimal value of $Z$, let us define the optimality factor $(C)$ as 
\begin{equation}
	C=\frac{Z}{Z_{opti}}.
\end{equation}

Consider the $(2,4N,N)$- Golay-ZCZ sequence sets described in Theorem \ref{th1}. If we consider the binary cases, then we have $C=1$. Hence the resultant Golay-ZCZ sequence pairs proposed in Theorem \ref{th1} are optimal.

Assuming that an $(M,M,N)$- CCC exists, from Theorem \ref{const2n}, we get $(M,M^2N,(M-1)N)$- Golay-ZCZ sequence set. To be an optimal Golay-ZCZ sequence set, the optimal ZCZ width $Z_{opti}=\frac{L}{M}$, i.e., in this case $Z_{opti}=MN$. However, the ZCZ we can achieve through Theorem \ref{const2n} is $Z=(M-1)N$.

Therefore,  
\begin{equation}
	\begin{split}
		C&=\frac{Z}{Z_{opti}}\\
		&=\frac{(M-1)}{M}. 
	\end{split}
\end{equation}
Therefore, $\lim\limits_{M\rightarrow \infty}C=1$, hence,
the sequence sets proposed in Theorem \ref{const2n} are asymptotically optimal, as the number of sequences increases.

\section{The novelty of the Proposed Constructions as Compared to Previous Works}\label{novel}
In this section we compare the proposed constructions to the previous works, specifically with the works of Gong \textit{et al.} \cite{Gong2013} and Chen \textit{et al.} \cite{Chen201811} and discuss the novelty of the proposed constructions. 
\begin{enumerate}
	\item In \cite{Gong2013}, the authors only considered the complementary sequences of lengths of the form $2^m$, and analysed their corresponding periodic zero auto-correlation zones. As compared to that, Theorem \ref{th1} and Theorem \ref{const2n} considers complementary sequences of a more flexible form and analyses both periodic zero auto-correlation zone as well as zero crosscorrelation zone.
	\item In \cite{Chen201811}, the authors proposed Golay-ZCZ sequence sets. However, the constituent sequences have lengths only of the form of $2^m$. As compared to that, in Theorem \ref{th1} we have proposed Golay-ZCZ sequence pairs of length $4N$, where $N$ is the length of a GCP. In Theorem \ref{const2n}, we have proposed Golay-ZCZ sequence sets, consisting sequences of lengths $M^2N$, using an $(M,M,N)$ CCC. As we see the lengths are more flexible as compared to the results in \cite{Chen201811}.
	\item The results reported in \cite{Chen201811} only achieves optimality for the case of binary $(2,2^m,2^{m-2})$- Golay-ZCZ sequence pairs. As compared to that, based on the discussions in Section \ref{secv}, we can see that Theorem \ref{th1} results to binary $(2,4N,N)$- Golay-ZCZ sequence pair, which is optimal. Theorem \ref{const2n} results to polyphase $(M,M^2N,(M-1)N)$ Golay-ZCZ sequence set, which is also asymptotically optimal, as the number of sequence increases.
	\item Analysing Table 1 of \cite{Chen201811}, we can observe that, as the number of sequences of the Golay-ZCZ sequence set increases, the ZCZ width decreases. As compared to that, in Theorem \ref{const2n}, as the number of sequences of the Golay-ZCZ sequence set increases, the ZCZ width also increases, and eventually we achieve asymptotic optimality.
	\item To increase the availability of CCCs of various parameters, which in turn increases the flexibility of the parameters of the proposed Golay-ZCZ sequence sets, we have proposed a new iterative construction of CCCs. We have also provided computer search results for binary CCCs with parameters $(4,4,N)$, for various values of $N$ given in Table \ref{tab3}. Moreover, using those as seed CCCs in \cite{jin} and Theorem \ref{th3}, we can construct several binary CCCs with new parameters, which are not reported before. As discussed in Remark \ref{remarkccc}, these CCCs are important in itself, since $N$ is not a power-of-two.
\end{enumerate}

\section{Conclusion}\label{section 5}

In this paper, we have made two contributions. Firstly, we have systematically constructed GCPs of lengths non-power-of-two where each of the constituent sequences have a periodic ZACZ and the pair have a periodic ZCCZ. We have also constructed complementary sets consisting of sequences having large periodic ZACZ and ZCCZ using CCCs and IDFT matrices. Notably, the second construction results to asymptotically optimal Golay-ZCZ sequences with respect to Tang-Fan-Matsufuji bound, as the number of sequence increases. To increase the availability of CCC for various parameters and consequently increase the flexibility of the proposed Golay-ZCZ sequence sets, we have also proposed a new iterative construction of CCCs. Moreover, we have found binary CCCs with parameters $(4,4,3)$, $(4,4,5)$, $(4,4,7)$, $(4,4,11)$, and $(4,4,13)$ based on computer search, which can be used as seed CCCs, and can hugely increase the flexibility of the proposed construction. Since the length of the resultant Golay-ZCZ sequences are more flexible, the proposed constructions partially fills the gap left by the previous remarkable works of Gong \textit{et al.} and Chen \textit{et al.}. The proposed Golay-ZCZ sequences have potential applications in uplink grant-free NOMA.



\end{document}